# Towards Blockchain for Decentralized Self-Organizing Wireless Networks

Steven Platt
*Department of Information and Communication Technologies Engineering*
*Pompeu Fabra University*
Barcelona, Spain
steven.platt@upf.edu

Miquel Oliver Riera
*Department of Information and Communication Technologies Engineering*
*Pompeu Fabra University*
Barcelona, Spain
miquel.oliver@upf.edu

*Abstract*— Distributed consensus mechanisms have been widely researched and made popular with a number of blockchain-based token applications, such as Bitcoin, and Ethereum. Although these general-purpose platforms have matured for scale and security, they are designed for human incentive and continue to require currency reward and contract functions that are not requisite in machine communications. Redes Chain is a custom designed blockchain, built to support fully decentralized self-organization in wireless networks - without a cryptocurrency or contract dependency.

## I. Introduction

Initially popularized through application in digital currency, distributed ledger technologies (DLT) are now seeing wider adoption as a path for extending peer-to-peer design and security to the broader internet. To allow open participation, a number of DLT designs deploy computationally expensive cryptography paired with digital currency reward [1], creating a format optimized for human incentive and trust that is a less natural fit for machines. Although popular applications of blockchain including Bitcoin and Ethereum, tie blockchain to a digital currency function - it is important to note that blockchain as a data structure has no native association to digital currency. It is this isolated application of the blockchain data structure, and its ability to support distributed consensus, that is the focus of this research.

The ability to form consensus among equal peers has a number of implications in networks research - but we propose its most natural application is that of decentralizing self-organization functions in wireless networks - systems which by design are dependent on coordination and context sharing among network participants.

In the following sections of this paper, we detail the use cases and limitations of existing cryptocurrency-based blockchain ecosystems Bitcoin and Ethereum. We then define the broader Self Organizing Network use case - presenting a new model, built atop blockchain, that is wholly decentralized. Next, we present Redes Chain - a custom blockchain prototype, designed for machine communications and to allow the decentralization of self-organization functions in wireless networks. Finally, the Redes blockchain prototype is demonstrated through a proof-of-concept deployment handling access federation among independent 802.11 networks.

## II. Bitcoin and Ethereum: Blockchain for Currency and Contracts

Today Bitcoin is the most well-known application of blockchain technology, but also the oldest and simplest technical implementation in popular use. The original Bitcoin whitepaper was published in 2008, and detailed the design of a digital currency system that removed the need for a trusted third party for the verification of transactions [1]. In being designed as a digital currency, the Bitcoin ledger structure can be simplified as a state transition system; with the current state represented as the total ownership of all digital coins at a moment in time, and the state transitions represented by the movement of these coins, or payments made between users in the network.

To aid in decentralization, Bitcoin is designed as a permissionless network, allowing nodes or participants to join and leave the network at any time - with all transaction data sent as broadcast. After data is broadcast, computers in the network compete to find a hash of the block data that is smaller than a threshold size set for the entire network. Because the result of hashing is pseudo-random, it is believed that every computer in the network of equal computing power, has an equal chance of being first to find the correct hash [2]. The equality created through the pseudo-random hash function also makes the network able to remain secure as long as a simple majority, or 51% of the network are acting in good faith. To provide incentive for doing the difficult computation work, machines participating in the network are issued a reward in the form of Bitcoin, for finding and broadcasting the first hash successfully meeting the size threshold. These reward payments are covered by transaction fees charged to users wishing to add blocks to Bitcoins' chain. Because this hash value can be verified by others in the network - this process of consensus is named "Proof of Work" (POW). A primary side effect of the race condition created through POW consensus, is that power used for all unsuccessful hashes is considered wasted, making the system highly resource inefficient. A secondary behavior and weakness of using the POW consensus model is that is makes financial incentive in the form of block rewards and transaction fees, native to the operation of the system.

In 2013, Vitalik Buterin published the Ethereum whitepaper, seeking to expand the functions of Bitcoin - into a general purpose compute platform. The Ethereum blockchain included a more complex block structure that allowed storing logic which executes only when preset

This work was partially supported by the Spanish and Catalan Governments through the project "Plan Nacional": AEI/FEDER TEC2016-79510 "Redes Con Celdas Densas y Masivas".

conditions are met. This new block structure allowed the creation of contracts, but retained Bitcoin's permissionless format, POW consensus, and block rewards [3]. Programmed into the contract support of Ethereum, is the ability to create a secondary digital currency token, pegged to the value of Ethereum's own digital currency, Ether. These tokens are referred to as ERC-20, and in effect, allow a white labeling of the core Ether token, while retaining compatibility with the broader ecosystem of Ethereum smart contracts [3]. A number of network systems have been built on top of the Ethereum blockchain, but in doing so, these systems must inherit Ethereum's contract structure, with peer-to-peer payments at the core. Examples of such systems includes Privatix Network, a VPN service allowing peer-to-peer payment for bandwidth used while hosting VPN connections [4], and Ammbr, a mesh networking service allowing peer-to-peer payment for users who agree to share their internet connection [5].

### III. SELF-ORGANIZING NETWORKS

To address concerns of increasing complexity in cellular networks, the 3GPP completed work to formally define Self Organizing Network (SON) functions in tandem with the development of the LTE cellular standard. SON functions today are formed in one of three designs: centralized, distributed, and hybrid [6]. In a centralized design, resource management and air interface coordination algorithms are processed by a central controller. With a distributed model, these algorithms are run at the network edge. Finally, a hybrid model employs a combination of the former two [6]. It is important to note that although a distributed model allows algorithms to run at the network edge, these controls remains limited to coordination within a single operator environment, with existing designs often relying on S2 interface connections to a carrier core in cellular deployment, or a hub controller in 802.11x networks for compatible hardware actuation and control [7].

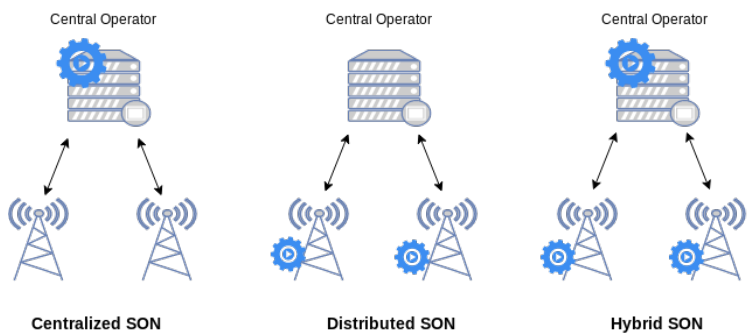

Fig. 1. SON Architectures

With a target to reduce manual administration by automating routine configurations in cellular networks; the SON standards developed by the 3GPP eventually included provisions for energy savings, handover optimization, automatic neighbor relation management, and load balancing [7]. Today these cellular-centric SON operations have also been extended to Wi-Fi and other air interface networks which benefit from the enhanced environment knowledge and distributed coordination capability that SON provides [8]. The Redes blockchain proof of concept, presents an example of wireless SON functionality, built in a new fully decentralized context, allowing wider coordination among previously isolated networks and nodes (fig. 1).

### IV. REDES: BLOCKCHAIN FOR DECENTRALIZED SELF-ORGANIZING WIRELESS NETWORKS

To allow operation in heterogeneous network environments, the Redes blockchain is structured as a permissioned chain and does not attempt to enforce code execution commitments in the form of smart contracts as with Ethereum. With Redes, data storage and hardware specific actuation and control operate separately (fig. 2). The Redes proof of concept as presented, deploys blockchain in its more basic form - as a decentralized data store for network specific state data, to support decentralized SON functionality in wireless networks. Structuring the chain in this way allows the benefits of developing fault tolerant state consensus without requiring central ownership, while leaving code execution control with individual network operators who can optimize for various combinations of infrastructure, that make use of the wider shared network context.

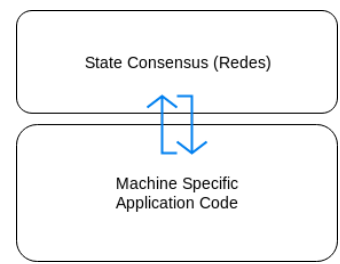

Fig. 2. Redes separation of concensus and application code

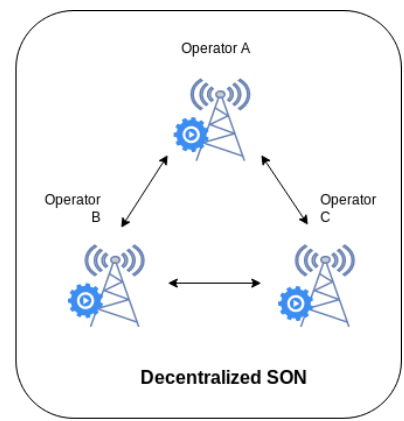

#### A. System Requirements

The Redes blockchain is written in the Python programming language, for hardware isolation, and portability across infrastructure. Using Python version three for the construction of Redes also allowed the use of Python libraries and micro-frameworks - including Flask and SQLite for full web server and database functions in a compact package for embedded systems use. Since it is not intended as a one-size application platform, or a digital currency, Redes can strip out dependencies that would require a full Linux operating system or more robust database systems as seen in larger Ethereum, and Bitcoin derived projects [3]. In current form, the Redes Blockchain code is 390 lines of Python code utilizing 13kb of disk space in isolation, and can be installed on feature restricted embedded operating systems, such as the 'Busy Box' Unix operating system, or any platform supporting Python version three.

*B. API, Block Format, and Concensus*

Making use of the Python 'Flask' micro-framework, Redes includes its own API with 6 initial functions: register a node, remove a node, trigger consensus, issue a transaction, create a block, and request the longest chain. Testing functionality and interacting with the underlying ledger is done through calls to these 6 API's.

Although the API format is consistent with other blockchain projects, the larger change is the format of the ledger blocks themselves. The block fields in Redes do not include provisions for currency, or account balance as in Bitcoin and Ethereum. These are replaced with 'mac address' and an 'action' field for SON control of network access in the proof of concept use (1). Table 1 shows a high-level comparison of Bitcoin, Ethereum, and prototype Redes blockchain.

$$def\ new\_transaction(self, sender, recipient, mac, action) \quad (1)$$

TABLE I. COMPARISON OF BITCOIN, ETHEREUM, AND REDES BLOCKCHAINS [9]

| Features | Blockchain System | | |
|---|---|---|---|
| | *Bitcoin* | *Ethereum* | *Redes* |
| Consensus | Proof-of-Work | Proof-of-Work | Modified Proof-of-Work |
| Participation | Permissionless | Permissionless | Permissioned |
| Language | C++ | Go, C++, Rust | Python |
| Currency | Bitcoin | Ether | - |
| Contract Support | Limited | Full | - |

For consensus, Redes uses a modified proof-of-work consensus algorithm. Proof of Work as implemented in Bitcoin and Ethereum, requires all network participants to compete in providing a successful hashing for new blocks, this is required in the permissionless format, where the contribution of blocks needs to be made random among open participants, to prevent influence and maintain security [1]. Being a permissioned chain, Redes removes this compute race condition, and requires the node issuing a transaction, to hash the block itself before broadcasting the valid hash. This update to the proof of work consensus makes computation and energy usage linear with transaction volume, rather than exponential as in Bitcoin and Ethereum. [1] Before a node accepts the transaction, the proof is still validated before it is added to a local chain. In this modified proof of work scenario, the integrity and consensus of state formed in the chain is still assumed valid, as any modification or corruption of previous block data changes and invalidates the hash result achieved by network nodes, when the proof is checked (fig. 3)[9].

Algorithm 1 shows the pseudocode logic behind the Redes modified POW consensus.

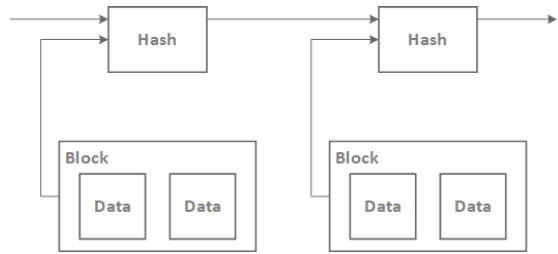

Fig. 3. Blockchain Data Structure [10]

Algorithm 1: Pseudocode for Redes Modified Proof-of-Work Consensus

```
1.  function resolve_conflicts(self)
            // constructor
2.          neighbors ← known nodes
3.          new_chain ← null
4.          old_chain ← local chain
5.          max_length ← block length of local chain

            // network synchronization
6.          for node ∈ neighbors
                    // request chain of all neighbors
7.                  response ← requests.get(https://{node}/chain)

8.                  if response = !null
9.                          length = response.length
10.                         chain = response.chain

                            // if chain is longer and also valid,
                            //replace the local chain
11.                 if length > max_length and self.valid_chain(chain)
12.                         max_length = length
13.                         new_chain = chain

14. function valid_chain(self, chain)
            // constructor
15.         last_block ← chain[0]
16.         currenct_index ← 1

17.         while current_index < length(chain)
18.                 block ← chain[current_index]

19.                 if block[hash] != self.hash(last_block)
20.                         return false

21.                 if !self.valid_proof(last_block[proof], block[proof])
22.                         return false

23.                 last_block = block
24.                 current_index + 1
25.                 process_son(block)
```

*C. Decentralized Network Access Control Use Case*

To test initial function of the Redes blockchain prototype, a testbed was devised, consisting of 3 802.11 capable wireless access points running the OpenWRT operating system, and installed inside VirtualBox. Running Redes within an installation of OpenWRT, the combined system, inclusive of the operating system, web server, and database - totaled less than 60Mb for its VirtualBox disk image. Other specifications for the OpenWRT hosts are 1 virtual CPU core, and 256Mb of RAM. The target of the testbed was to prove an early application of Redes state consensus, combined with local application control to execute the decentralized SON function of network access control among the otherwise isolated wireless access points.

Beginning with the base OpenWRT image, the three systems had all dependencies installed, then set to run the Redes blockchain. After this initial validation, the Redes blockchain API was validated using the 'Postman' API testing application. All nodes were registered with each other, using the 'register a node' API function, to allow syncing and writing the Redes blockchain (figure 4). Each OpenWRT system then issued 'create a block' transactions to the Redes JSON API interface, with a value filled for 'mac' and an additional 'action' field, signaling a device should be 'allowed' or 'denied' network access (figure 5). In total, validation of the 6 API functions was successful, proving that the desired data was stored in the chain and consensus based on the longest chain could be formed - although network access was not yet tied to information in the chain.

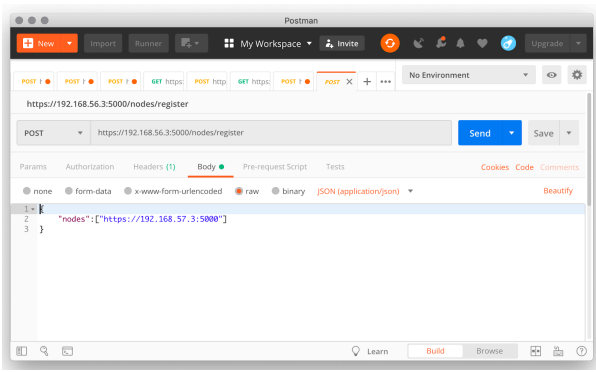

Fig. 4. New node registration using the Postman utility and Redes JSON API

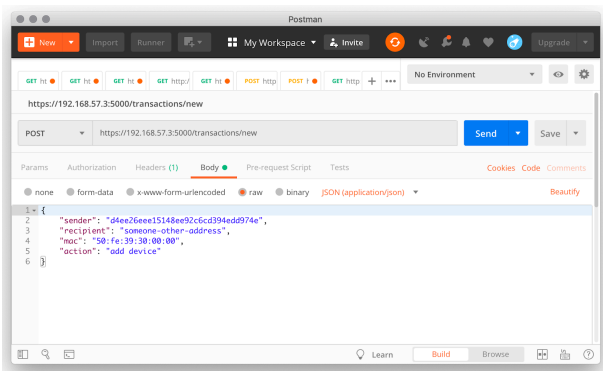

Fig. 5. Issuing a new block using the Postman utility and Redes JSON API

A final test from the testbed required pairing local code execution with the Redes state consensus, allowing the OpenWRT systems to issue system commands to add and remove devices from its local firewall configuration - allowing access to the previously isolated wireless networks (algorithm 2).

Algorithm 2: Pseudocode for SON Local Execution Within OpenWRT

1. function process_son(block)
       // constructor
2.     device_mac ← block(mac)
3.     device_action ← block(action)

4.     if device_action = add
           // openwrt platform specific commands
5.         openwrt subprocess.call('add firewall rule [device_mac]')
           // apply configuration to the last firewall rule added
6.         openwrt subprocess.call('set firewall.rule[-1].target=accept')
7.         openwrt subprocess.call('set firewall.rule[-1].proto=tcp udp icmp')
8.         openwrt subprocess.call('set firewall.rule[-1].src=lan')
9.         openwrt subprocess.call('set firewall.rule[-1].src_mac=[device_mac]')
10.        openwrt subprocess.call('commit and reload')

11.    if device_action = remove
12.        openwrt subprocess.call('delete firewall rule [device_mac]')
13.        openwrt subprocess.call('commit and reload')

Running the updated Redes code successfully allowed adding new blocks, forming consensus based on the longest valid chain, and finally adding the new devices to local firewall rules, demonstrating a decentralized SON use case of federating access controls to the additional OpenWRT devices operating the Redes blockchain.

V. DISCUSSION AND FUTURE WORK

In this research, we outline popular applications of the blockchain data structure and its historical application and limitation to currency and contract use. Next, we outlining how a modification of blockchain technologies, to remove currency and contract dependencies, could provide a better fit for machine communication, and specifically the full decentralization of SON functions in wireless networks. Finally, we demonstrate the Redes blockchain, a custom blockchain, developed using the Python programming language to allow isolation from hardware and wide compatibility, to handle the proposed decentralized SON function.

In current form, the Redes blockchain prototype is limited. The testbed delivers a basic permissioned blockchain that can be used to share and action data, while consuming less resources in embedded wireless network application, through a modification of classic proof of work algorithms. Redes validates a path of use-case specific blockchain vs use of existing chain ecosystems for our wireless network SON use case, but has not been fully stress tested or designed for scale. Future research for Redes is planned to focus on development of its consensus model, and support for a more standardized and complete suite of wireless network SON functions, such as neighbor discovery, power control, and wireless channel selection.